\documentstyle[12pt,psfig,aaspp4]{article}

\input epsf.tex

\newcommand{\be}{\begin{equation}}
\newcommand{\ee}{\end{equation}}
\newcommand{\nn}{\mbox{} \nonumber \\ \mbox{} }
\newcommand{\ba}{\begin{eqnarray}}
\newcommand{\ea}{\end{eqnarray}}

\newcommand{\Alfven}{ Alfv\'{e}n }

\begin{document}

\title{ Mass loading of pulsar winds
}

\author{
M. Lyutikov}
\affil{Department of Physics, McGill University,
Montr\'eal, QC \\
Massachusetts Institute of Technology,
77 Massachusetts Avenue, Cambridge, MA 02139\\
CITA National Fellow}

\begin{abstract}
The dynamics of relativistic magnetized mass loaded 
outflows carrying toroidal magnetic field  is analyzed in the context
 of Pulsar Wind Nebulae (PWNs).
Mass loading is very efficient in slowing down  super-relativistic
magnetized  flows and weakening of relativistic
shocks. We suggest that weakening of relativistic
reverse shocks by mass loading in PWNs is responsible
for the low radiative
 efficiencies of the majority of the PWNs.
Mass loading 
  may  also  result in a  shock transition near the 
 fast magnetosonic point; this is unlikely to happen in 
majority of PWNs.
The evolution of  magnetized mass loaded
flows beyond the reverse shock
is complicated: after initial deceleration to the minimal
velocity required to transport the magnetic flux,  the mass loaded flows 
 have  to  {\it accelerate}.
In order to be able to expand to infinity, magnetized flows should
either become time dependent or 
destroy the toroidal magnetic flux 
 by developing internal instabilities.
Destruction of the magnetic flux initiated by mass loading 
may allow for the flow to slow down to sub-relativistic velocities
and resolve the $\sigma$ paradox of the pulsar wind nebula.
\end{abstract}

\section{Introduction}

In this paper we explore the dynamics of  magnetized relativistic
pulsar winds subject to mass loading.
We are motivated by several possible applications: (i) slowing down of the
high-$\sigma$ flows ($\sigma$ is conventionally defined as a ratio
 of Poynting to particle fluxes)
 inside the  pressure confined
 pulsar wind nebulae  (referred below to as a ``$\sigma$ paradox''),
(ii) low X-ray efficiencies and a conspicuous lack of radio PWN
around many energetic pulsars
and (iii) the structure
of the ram pressure confined pulsar wind nebulae.

{\bf (i)}.
The $\sigma$ paradox is a  long standing problem in pulsar physics 
(Rees \& Gunn 1974, Kennel \& Coroniti 1984). Models of the pulsar 
magnetosphere (Goldreich \& Julian 1969,
 Arons \& Scharlemann 1979, Ruderman \& Sathurland 1975) 
 predict that  near the light cylinder  most of the spin-down luminosity of  
a pulsar 
 should be in a form of Poynting flux, $\sigma \gg 1$.
On the other hand,  modeling of the dynamics of the Crab  nebula
gives a low value of $\sigma$ at what is commonly believed to be 
a  reverse shock - strongly
magnetized  flows cannot  match boundary conditions
 (Kennel \& Coroniti 1984).
Several possibilities have been proposed to resolve the $\sigma$ paradox.
 One possibility is that $\sigma$ may change  between the 
 light cylinder and the reverse shock (Michel 1994, Coroniti 1993, Melatos \& Merose 1996,
critique by  Lyubarsky \& Kirk 2001).  
Alternatively, $\sigma$ may
  change  between the reverse shock and the contact discontinuity.
Begelman (1998) argued
that the post-shock flow may be distorted by  kink instabilities
which would annihilate the magnetic flux and allow strongly magnetized
flows to match  the boundary conditions at the edge of the nebula
(see also critique by Arons 1998).
Arons (2001) has suggested that interaction of the wind with the
ejecta filaments  may slow down the wind. In this paper
we investigate the latter possibility.

{\bf (ii)}.
Most PWNs, 
with the  notorious exception of the  Crab,  have very low 
efficiencies for the conversion of the pulsar spin-down luminosity
$\dot{E}_0$ into  X-rays, $L_X \sim 10^{-3} \dot{E}_0$ 
 (Seward  \& Wang 1988,  Becker \& Truemper 1997). No simple combination
of the pulsar parameters (like period and period derivative) 
can be fitted to the X-ray luminosity, which suggests
that environmental effects may be important (Possenti et al. 2001). Similarly, 
sensitive radio searches have failed to detect
PWNs towards selected energetic pulsars  (Gaensler et al. 2000).
 We show that mass loading of the pulsar wind
is very effective in slowing down the wind and thus weakening the 
termination shock. This may be  responsible for 
low X-ray and radio efficiencies of the PWNs.

{\bf (iii)}
In the past several years a new well defined class of objects - ram pressure
confined pulsar wind nebulae - has been established. Apart from a simple 
 model by Wang et al. (1993) no
realistic model has been proposed  to account for the often unusual structure 
structure of such PWNs. Bucciantini \& Bandiera (2001) have  argued that
classical bow-shock models cannot be applied to the  ram pressure
confined PWNs and suggested that mass
loading may be important.

The importance of mass loading on the pulsar wind  may be  estimated
using the Crab nebula as an example. It has long been known that 
 that some thermal plasma does evaporate off the trapped ejecta 
(Wilson 1974, Michel et al. 1991). It is only natural to expect that the 
coupling of a ``light'' pair plasma by  the ``heavy'' 
particles from the ejecta would strongly affects the dynamics of the
light fluid.
Recent reevaluation of the mass of Crab filaments (Fesen 1997) gives
$\sim 4.6 \pm 1.8 M_{\odot}$, while the total mass ejected into the
Crab nebula by the pulsar over its lifetime is tiny
$\sim 10^{-12}
M_{\odot}$ (at a rate $ \dot{N}_0 = 10^{38}$ pairs/sec
 (e.g., Kennel \& Coroniti 1984);
 ions may have a two order of magnitude larger mass flux; Galant et. al 1992).
Below we show that
mass loading starts to strongly affect the dynamics of  sub-sonic flows
when the pick-up mass flux, $\dot{M}$
 becomes of the
order of the total energy flux $\dot{E}_0$  divided by $c^2$ (eq. (\ref{qp})).
The total  pick-up mass flux is 
 $ \dot{M} \sim M_{\rm picked} v /r$ 
(with velocity given by the expansion velocity $v \sim 2000$ km/sec  
of the Crab nebula with the radius $r\sim 2$ pc; Kennel \& Coroniti 1984).
  The  mass
flux  required to slow down the flow to match the boundary
$M_{\rm picked} \sim \dot{E}_0 r /( c^2 \, v)  = 3 \times 
10^{27} g  $, a tiny fraction
of the available mass.   
 Thus
even a weak coupling of  filaments to the flow may strongly affect its
evolution. This may be contrasted with the theoretical estimates
of   Balbus (1981) who have 
 argued that filament evaporation
by conduction  may occur on a very  short time scale, of the order of the
dynamical time scale.

In the ram pressure confined PWNs, we should substitute for $v$ the velocity
of the pulsar $v \sim 200 $ km/sec and for $r$ the stand-off distance, which
is typically a fraction of a parsec.
The  total  pick-up mass flux, $ M_{\rm picked} \sim 10^{24}$ g,  is 
nevertheless smaller since typical pulsars 
have $\dot{E}_0 \sim 10^{35}$ erg/sec. This is approximately the
amount of the ISM material contained in the head part of the
ram pressure confined PWN.

\section{Model assumptions}

We  assume that an interaction between the trapped ejecta and the pulsar winds
occurs through mass loading of the  wind.
Filament material may be coupled to the pulsar wind by thermal evaporation,
hydrodynamic ablation and by photo- and electro-evaporation by high energy
photons and relativistic electrons. Interaction of the ISM clumps with 
non-relativistic winds have been studied  extensively
(Borkowski et al. 1990,  Cowie \& McKee 1977,  McKee \& Cowie  1977, 
Hartquist et al. 1986). No treatment of the interaction of the
ISM material with relativistic magnetized  flow exist; 
estimates based on the models of suprathermal evaporation 
(Balbus \& McKee 1982) predict a very high rate (Balbus 1981). 
Here we are interested in the dynamical effects of the mass loading
on the relativistic flow; 
hence we do not consider here the 
microphysics of the flow-filament interaction 
and instead we parameterize it
by a capture rate, (eq. (\ref{x3})),
 taken to
be  a prescribed function of a distance from a pulsar.
 The newly acquired particles are
assumed to be immediately coupled to the flow. \footnote{Relative streaming
of the newly ionized component with respect to the bulk flow may result in 
ion-cyclotron instabilities and turbulence generation (e.g., Galeev et al. 1996)}
We neglect the energy losses required for the 
possible ionization  and
the energy associated with the thermal motion of donor particles.
Mass loading inside a shock may also be important (Zank \&
Oughton 1991). Mass loading of relativistic shocks is briefly considered in
the Appendix \ref{shock}.

The dynamics of the mass loaded flows turns out to be non trivial
and somewhat counterintuitive. The overall dynamics of the mass loaded
flows resembles a nozzle-type flow where 
the rate of mass-loading contributes a
 term which  may be considered as a negative
pressure (eq. (\ref{dg})). Hence 
the effects of mass loading 
are opposite to that of pressure.
For example, 
 contrary to the
naive guess, the loading of strongly magnetized pulsar
wind leads not to slowing down
 but to {\em acceleration } of the flow. 
Acceleration formally proceeds to arbitrary larger velocities which, naturally,
cannot be realized in reality.
We argue that such 
acceleration of the wind will result in development of internal
 instabilities (which we have not addressed  yet)  that
would destroy the toroidal
 magnetic flux and, in a way similar to suggestion by  Begelman (1998),
would eventually allow a wind to slow down.

  The pulsar wind model
that we adopt is similar to the Kennel \& Coroniti model:
 spherical outflow of a polytropic gas
 with a toroidal magnetic field. In addition, the wind is assumed to
be  loaded by ejecta  
particles which are initially at  rest. 
By neglecting the radial component of the magnetic field
 we limit the applicability of this approach to
asymptotically large distances, far from 
the acceleration region of the flow (Weber \& Davis 1967, 
Goldreich \& Julian 1970,
Kennel et al. 1983). 
Typically the launching of the  pulsar wind is thought to occur
close to the light cylinder, with the flow reaching
 super fastmagnetosonic
velocities by passing successfully  through the slow, \Alfven
and fast  sonic  points (Goldreich \& Julian 1970).
At  distances large compared with the  light cylinder radius,
the flow usually can be well  approximated
as a radial motion with toroidal magnetic field,
 launched
with some given parameters: mass, energy and magnetic fluxes
loss rates.

Pulsar winds are launched at super-fastmagnetosonic velocities and have
to decelerate in a shock transition in order
to be matched onto the 
non-relativisticly moving external media (the SNR ejecta or ISM).
Under the hydrodynamic
approximation, a contact discontinuity separates  the
shocked wind material from the external medium. In some circumstances, the
external material
may get inside the contact discontinuity and even inside the reverse
shock propagating in the wind. This may be due to
either  incomplete ISM or ejecta ionization (neutral particles  may not  be
well coupled to the ionized component on time scales of interest to us)
or by evaporation of the ISM or ejecta  clumps trapped by the expanding wind.
Coupling of the newly added 
 material to the flow may considerable change its evolution.

The evolution of non-relativistic mass-loaded flows has been investigated
in a number of papers, starting with the Solar wind-comet
interaction (Biermann et al. 1967). Lately Hartquist et al. (1986), Arthur et al. (1993), Smith (1996),
Williams et al. (1999), Toniazzo (2001)  have
  investigated effects of mass loading on 
stellar outflows. Here we   show that  the 
 presence of a magnetic field 
makes it formally impossible for the mass loaded  flow to reach infinity.
In order to be able to reach infinity,
 the flow must either become non-stationary
or destroy its
magnetic flux, presumably by developing internal instabilities
which result in reconnection.

\section{ Governing Equations}

The formal treatment of the problem starts with the 
set of relativistic magnetohydrodynamic equations which 
can be written in terms of conservation laws (e.g. Landau \& Lifshitz 1975):
\ba && 
T^{ij}_{,i} =S_i,
\label{x1}
 \\ &&
F^{\ast\, ij}_{,i} =0,
\label{x2}
\\ &&
(\rho u^i)_{,i}=R
\label{x3}
\ea
where
\be
T^{ij} = (w +b^2) u^i u^j +(p+{1\over 2} b^2) g^{ij} -b^i\,b^j
\ee
is the stress-energy tensor,
$w =\rho+ {\Gamma \over \Gamma -1} p$ is  the 
 plasma proper enthalpy, $\rho$ is proper plasma density and $p$ is pressure,
$b^2 = b_i b^j$ 
is the plasma proper magnetic energy density times $4 \pi$, 
$p$ is pressure, $u^i=(\gamma, \gamma {\bf \beta})$ 
are the plasma four-velocity, 
Lorentz-factor and three-velocity, $g^{ij}$ is the metric tensor, 
$b_i = {1\over 2} \eta_{ijkl} u^j F^{kl}$ are the four-vector of magnetic
field, Levy-Chevita tensor and electro-magnetic field tensor,
 $S_i$ is a four-vector representing energy and momentum sources
and $R$ is the density source
 (dimensions of R are ${\rm g\, cm^{-3}\, sec^{-1}}$). 
We assume that the adiabatic index $\Gamma$
 is constant for algebraic simplicity.
Since we are interested in transonic transitions for relativistic fluid
motion, we expect that a relativisticly hot flow with $p \gg \rho$ will have
an adiabatic index of $4/3$  while strongly subsonic (non-relativistic)
 flows will asymptote to
$\Gamma=5/3$.

We assume that loading is due to the medium at rest in the laboratory frame,
$ {\bf S}=0$, $S_0 = R$.
\footnote{
The
filaments are moving with velocities up to 300 km/sec 
(Trimble 1968), much smaller than the  velocity of the 
pulsar wind near the termination shock $v_{\rm shock}\sim c$ or near the
edge of the remnant $v \sim 2000$ km/sec (Kennel \& Coroniti 1984).
}
Writing  out eqns. (\ref{x1}-\ref{x3}) in coordinate form and 
assuming a stationary,
 spherically symmetric outflow with toroidal magnetic field,
we find
\ba 
&&
{1\over r^2} \partial_r \left[ r^2 ( w+b^2) \beta \gamma^2 \right] =R
\label{x4} \\ &&
{1\over r^2} \partial_r 
\left[r^2\left((w+b^2) \beta^2 \gamma^2 + (p+b^2/2)\right) \right] - 
{2 p \over r}=0
\label{x5} \\ &&
{1\over r} \partial_r  \left[r b \beta \gamma \right] =0
\label{x51} \\ &&
{1\over r^2 } \partial_r  \left[r^2 \rho  \beta \gamma \right] =R
\label{x6}
\ea
The above relations can be simplified if one 
introduces energy, mass and  magnetic flux loss rates,
\ba && 
{\cal L} = \beta \,{\gamma }^2\,\left( b^2 + {\Gamma \over \Gamma -1} 
 \,p + \rho  \right) \\ &&
{\cal F} = \beta \,\gamma \,\rho \\ &&
{\cal K}=  \beta \,\gamma \, b
\ea
which evolve according to equations
\ba && 
\partial_r {\cal L} = - {2 {\cal L} \over r} + R \\ &&
\partial_r {\cal F} = - {2 {\cal F} \over r} + R \\ &&
\partial_r {\cal K } = - {2 {\cal K} \over r} 
\ea
 which have solutions
\ba &&
{\cal F} = {\dot{M}_0 + \dot{M}\over 4\,\pi \,r^2} 
\nn &&
{\cal L } = { \dot{E}_0+ \dot{M} \over 4\,\pi \,r^2} 
\nn &&
{\cal K} =  { {\cal E} \over   2\, \sqrt{\pi }  r}
\ea
where
$\dot{M}_0$ and  $\dot{E}_0$ are the central source's mass and 
 energy  loss rates, ${\cal E}$ is the electromotive force
 and we have introduced
\be
\dot{M} = 4 \pi \int R\, r^2 dr
\ee
for the acquired mass flux.

It is convenient to introduce two other parameters of the flow:
magnetization $\sigma$ and a fast magnetosonic  wave  phase velocity
$\beta_f$ and Lorentz
factor $\gamma_f$
\ba &&
\sigma={ b^2 \over w} = {{\cal K}^2  \over {\cal L } \beta - {\cal K}^2}=
{{\cal E}^2 \over \dot{E}_0 \beta - {\cal E}^2}
\nn &&
\beta_f^2=  
{\sigma \over 1+ \sigma} + {\Gamma p \over (1+ \sigma) w}
\nn &&
\gamma_f^2 \equiv {1 \over 1- \beta_f^2}= 
{ 3   {\cal L } (1+\sigma) \over 2  {\cal L } + {\cal F} \gamma  (1+\sigma)}
\label{pp}
\ea

In particular for a relativisticly hot plasma, $\rho \ll p$, 
we have $w = {\Gamma  p /( \Gamma -1)} $  and
\be
\beta_f^2= { \sigma + \Gamma -1 \over 1+ \sigma} = 
{1 +  3 \sigma \over 3 ( 1+ \sigma)} \mbox{ for $\Gamma =4/3$.}
\ee
We will need also an expression for pressure in terms of the fluxes:
\be
p= (\Gamma-1)\,{ {\cal L } - {\cal F}\,\gamma  ( 1+\sigma)
 \over \Gamma \beta \gamma^2 ( 1+\sigma)} =
\frac{\left( \Gamma -1  \right) \,
    \left( {\cal L } - {\cal F}\,\gamma \,{{{\gamma }_f}}^2 \right) }{\beta \,
    {\gamma }^2\,\left( 2 - \Gamma  \right) \,\Gamma \,{{{\gamma }_f}}^2}
\ee
Since $p>0 $ it follows that  $ {\cal L } > \gamma\, (1+\sigma)\,{\cal F} $
and ${\cal L } >  \gamma\,  \gamma_f^2	\, {\cal F}$.

Eliminating 
${\cal K}$ in favor of $\gamma_f$ we 
get a particularly transparent form for the evolution of Lorentz
factor
\be
\left( \frac{\left( 2 - \Gamma  \right) \,
      }{
      \beta^2 \,   {\gamma }^3  \,{\left( \Gamma-1  \right) }}
    \right) \, 	\left( {\gamma }^2 - {{{\gamma }_f}}^2 \right) \partial_r \gamma =
\frac{2\,\left( {\cal L } - {\cal F} \,\gamma \,{{{\gamma }_f}}^2 \right) }
  {{\cal L}\,r } -
\frac{\left( \gamma-1  \right) \,\left( 2 - \Gamma  \right) \,
    \left( 1 + \gamma \,\Gamma  \right) \,{{{\gamma }_f}}^2}{{\cal L } \,
    {\left( \Gamma-1  \right) }}\, R 
\label{dg}
\ee
Equation (\ref{dg}) has a form of  nozzle-type  hydrodynamical flows
(e.g., Landau \& Lifshitz 1999); in astrophysical context it is best known
for Parker's solutions of the solar wind (Parker 1960). 
The lhs of  eq. (\ref{dg}) contains a familiar 
special point at the sonic transition $\gamma= \gamma_f$.
The positively defined first term on the rhs  describes the  evolution
of Lorentz factors due to pressure effects, and
the negatively defined second term is due to mass loading.
Thus, the rate of mass loading may be considered as a negative 
pressure.

By neglecting the radial magnetic field far from the acceleration
region we have lost the magnetic sling-shot
effect often evoked for the acceleration of plasma
(Michel 1969, Goldreich \& Julian 1970, Kennel et. al  1983). 
The upshot of those 
works is that the flow is accelerated to  supersonic velocities with
a terminal Lorentz factor $\gamma \sim \sqrt{\sigma}$. 
Since it is believed that the plasma near the light cylinder has
$\sigma_0 \sim 10^3-10^6$, but the inferred pulsar 
wind Lorentz factor is $\sim 10^6 \gg \sqrt{\sigma_0}$,
 an additional acceleration is 
required.

Without mass loading the rhs of eq. (\ref{dg}) is always
positive so that  super-fastmagnetosonic flows accelerate while
sub-fastmagnetosonic flows
 decelerate. With mass loading the situation is more complicated. 
Mass loading enters in the eq. (\ref{dg}) in two ways:
explicitly via the rate of mass loading $R$, given by the
 negatively defined second term on the rhs, and implicitly via
the accumulated mass and energy  fluxes in ${\cal F}$ and  ${\cal L}$.
The  negatively defined second term could in principle become
 larger than the pressure term. Below we show that this indeed happens
for magnetized flows.

A quick examination of eq. (\ref{dg}) reveals an important fact:
for strongly relativistic flows 
the mass loading term is enhanced by a factor $\sim \gamma^2$. 
For typical pulsar winds $\gamma \sim 10^6$, thus 
 mass loading is  extremely efficient in slowing down the wind.
We defer a detail examination of the eq (\ref{dg})
until  Section \ref{evol} and review  beforehand the evolution of 
unloaded flows.

\section{Dynamics of relativistic flows without mass loading}

To guide us through the effects of mass-loading we first derive
 the relations governing the evolution
 of unloaded relativistic magnetized winds.
The properties of these solutions have been discussed extensively
(e.g. Kennel \& Coroniti 1984), but the exact analytical form given below
has not been written, according to our knowledge.
First, we
introduce two new parameters (instead of $\dot{M}_0$
and  ${\cal E}$) - 
 the unloaded wind terminal Lorentz factor $\gamma_0$ and the
 initial magnetization parameter $\sigma_0$ defined by the following reactions
\ba &&
\gamma_0 = { \dot{E}_0 \over \dot{M}_0 (1+ \sigma_0)},
\nn &&
\sigma_0 = { {\cal E}^2 \over \gamma_0 \beta_0 \dot{M}_0}.
\ea

\subsection{Unmagnetized unloaded  relativistic flows}

In the absence of a magnetic field 
 the fast magnetosound  velocity is
\be
\beta_f^2 = \left( \Gamma -1  \right)
\frac{\left(  {\cal L}  - {\cal F} \,\gamma  \right)  }{{\cal L}}.
\ee
In this case 
 the equation for the evolution of the 
wind Lorentz factor (\ref{dg})
 can be integrated exactly:
\be
r \propto{1 \over \sqrt{\beta}} \, { (\gamma_0 - \gamma)^{1/(2( \Gamma -1))} \,
\gamma^{ ( 2 -\Gamma)/(2( \Gamma -1))} }.
\label{rg}
\ee
This  shows that  there are two branches of  solutions:
supersonic   and subsonic.
The supersonic branch
 after initial acceleration with
$\gamma \sim r^{2( \Gamma -1) /( 2 -\Gamma) }  $
($ \gamma \sim r$ for $\Gamma =4/3$)  
 reaches a terminal Lorentz factor
$\gamma_0 = \dot{E}_0 / \dot{M}_0$ (with $\gamma_0 - \gamma \sim 
r^{- 2( \Gamma -1)}$).
In the accelerating part the flow has $p \gg \rho$ - it is a pressure
driven acceleration. Since
pressure 
falls off with distance faster than density (for $\Gamma >1$),
 a transition  to the coasting phase with $\gamma \sim \gamma_0$
occurs at $p \sim \rho$.
 The subsonic branch
decelerates  to zero velocity at infinity with $ \beta \sim 1/r^2$, 
 keeping  pressure
and density almost constant, determined by the mass conservation ratio.

\subsection{ Relativistic Magnetized Unloaded Flows}

 Magnetized flows also have two  
 branches: subsonic and supersonic.
 The  terminal velocity is
 determined from the condition
$\partial_r \gamma =0 =p$, which, using eq. (\ref{pp}), can be written
\be
{ \gamma - \gamma_0 \over \gamma_0} =
{ \beta - \beta_0 \over \beta} \sigma_0.
\label{wp}
\ee
Eq. (\ref{wp})
 generally has two solutions: a  supersonic one
\be 
\gamma = \gamma_0 
\ee 
(it is highly supersonic for $  \gamma_0  \gg \sqrt{\sigma_0}$)
 and
a subsonic one, which in the limit $\gamma_0 \gg \sqrt{\sigma_0}$ gives
\be 
\beta = { \sigma_0 \over 1+ \sigma_0 }.
\label{sw}
\ee
No physical solutions exist for $\beta < \sigma_0 / (1+ \sigma_0)$ -
such flows
 cannot transport enough  magnetic flux.

The supersonic solution behaves similarly to the unmagnetized case.
The flow is accelerated by pressure effects as long as $p\gg \rho$, reaching
a  coasting phase  with $\gamma \sim \gamma_0$ when $p \leq \rho$.
Behavior of the subsonic branch is qualitatively different from the 
unmagnetized case. To study the behavior of the  subsonic branch 
we can use   a simplifying assumption
 $\gamma_0 \rightarrow \infty$  
since  for subsonic flows the Lorentz factor 
is usually only weakly relativistic,
 $\gamma \leq \sqrt{( 1 + \sigma_0)/(2 -\Gamma)}$ 
 while  $\gamma_0 \gg \sqrt{\sigma_0}$.
 The limit  $\gamma_0 \rightarrow \infty$ 
is equivalent to neglecting the mass loss rate of the central
source if compared with
the energy loss rate. 
The evolution of the flow is then given by
\be
r\propto
  \left(  \beta  \gamma \right)^{(2-\Gamma)/(2(\Gamma-1))} 
\left(  \beta  - \sigma_0 /(1+ \sigma_0 ) \right)^{-1/(2(\Gamma-1))}
=
 { \beta  \gamma \over (\beta - \sigma_0 /(1+ \sigma_0 ))^{3/2}}
\, \mbox{for $\Gamma=4/3$}.
\ee
This shows that  subsonic flows reach a minimum velocity  given by eq. 
(\ref{sw}).
 The fact that 
 magnetized flows cannot slow down to zero velocity since they have to
transport magnetic flux   is crucial in determining the asymptotic
dynamics of the mass loaded flows.

\section{Evolution of mass loaded pulsar winds}
\label{evol}

A pulsar produces  a wind with properties determined by three
parameters: energy  flux $\dot{E}_0$,  terminal Lorentz factor $\gamma_0$
and magnetization $\sigma_0$.
 We also assume that  near the acceleration region
(located presumably close to the light cylinder)
the mass loading is unimportant, so that  the wind has reached
a strongly super-fastmagnetosonic 
 terminal  Lorentz factor  $\gamma_0  \gg \sqrt{1+ \sigma_0}$
(this requires 
$\dot{M}_0 \ll {\cal E}^2 /\sigma_0, \, \dot{E}_0 \beta_0 /(1+\sigma_0)
$) and its internal pressure
has dropped to 0.

The ratio of the mass flux due to loading to  the
initial energy flux is 
\be
\zeta = { 4  \pi \int  R  r^2 dr \over   \dot{E}_0} =
{ 4  \pi R r^3 \over (3-n ) \dot{E}_0}
\label{d1}
\ee
for a power-law  dependence of mass loading  on radius,
 $R \sim r^{-n}$.
Parameter $\zeta$ will often be used instead of the radius $r$.

\subsection{
Weakly loaded flow}

Here we show that mass loading of relativistic flows
is extremely efficient. Consider evolution of a strongly
super-fastmagnetosonic cold flow when mass loading may be approximated
as a small perturbation.
In the limit $ r \rightarrow 0$ and $\gamma, \, \gamma_0 \gg 1$,
eq. (\ref{dg}) gives
\be
{ \partial \ln \gamma  \over  \partial \ln r} 
= { 2(\gamma_0  - \gamma) \over  \gamma + (2 - \Gamma) \gamma_0 /(\Gamma -1)}
- { 4  \pi r^2 R \over  \dot{E}_0 } 
\, 
{ \Gamma \gamma^2 \gamma_0 (1 + \sigma_0) \over \gamma (\Gamma -1)  + (2 - \Gamma) \gamma_0 }.
\ee
If initially $\gamma = \gamma_0$, then
\be
{ \partial \ln \gamma  \over  \partial \ln r} 
= - { 4  \pi r^3 R \over  \dot{E}_0 } \Gamma \gamma_0^2 (1+\sigma_0)
= - { 4  \pi r^3 R \over  \dot{M}_0 } \Gamma \gamma_0 
\ee

Thus, the flow starts to decelerate quickly when
the accumulated mass flux becomes comparable to $ \dot{M}_0  / \gamma_0$ - 
an increase in efficiency of loading by a factor of $\gamma_0$ 
($\sim 10^6$!).
The typical 
scale for the deceleration of the flow is 
\be
r_d \sim \left( {\dot{E}_0  \over 
4  \pi R \Gamma \gamma_0^2 (1+\sigma))} \right)^{1/3}=
 \left( {\dot{M}_0  \over 4  \pi R \Gamma \gamma_0}\right)^{1/3}
\label{r_d}
\ee
which is $\gamma_0^{1/3}$ times smaller than in the non-relativistic case.

Efficient deceleration of supersonic flows by mass loading come from the
requirement that the newly acquired particles, which were initially at rest,
have to be accelerated to large Lorentz factors. This costs a lot of energy
and momentum.

\subsection{Special points in the flow}

Special point in the  flow occur
when the rhs of the eq. (\ref{dg})  becomes 0.
It is shown below that 
 in  unmagnetized flows, the only
special point  in the flow  occurs
where $\beta=\beta_f$, in order to keep $\partial_r \beta$ finite, while
mass loading of magnetized flows introduces another special point
where $\beta \neq \beta_f$ and, thus, 
$\partial_r \beta  = \infty$.

\begin{figure}
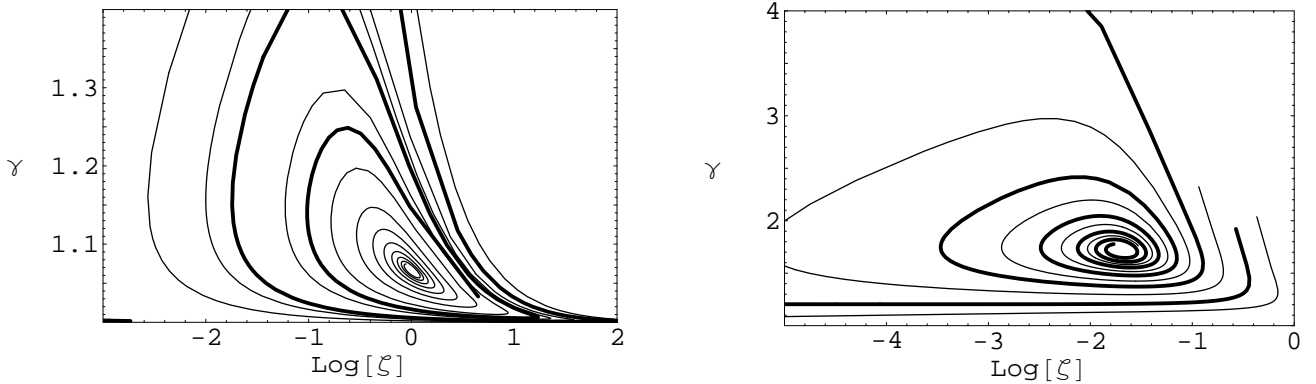

\centerline{
\psfig{file=foc-relnonma.eps,width=9cm}
\psfig{file=foc-relma.eps,width=9cm}}
\caption{ Phase portrait  of relativistic magnetized mass loaded flows
($n=0$, $\Gamma=4/3$).
{\bf Left:} Unmagnetized case $\sigma_0=0$.
{\bf Right:} Magnetized flow with $\sigma_0=1$.
Thick lines are the critical solutions which start at 
$\gamma=\gamma_0$ and $\beta= \sigma_0/(1+ \sigma_0)$.
}
\label{toporel}
\end{figure}

\begin{figure}
\centerline{
\psfig{file=focus-nonma.eps,width=8cm}
\psfig{file=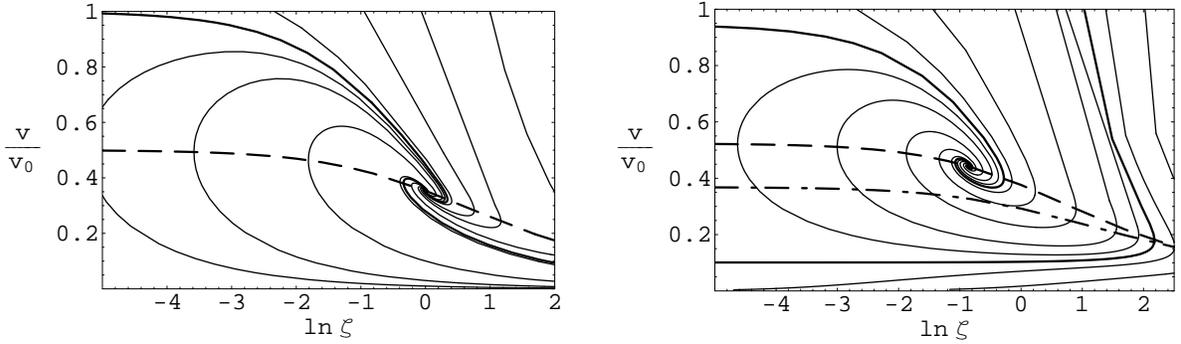,width=8cm}}
\caption{  Phase portrait  of  non-relativistic 
mass loaded flows ($n=0$, $\Gamma =5/3$).
{\bf Left:} Unmagnetized case $\sigma_0=0$; focus is located at
$\zeta=1$, $v=v_0/\sqrt{8}$ (compare with Smith 1996).
{\bf Right:} Magnetized flow with $\sigma_0=0.1$; focus is located at
$\zeta=-.86$, $v= 0.44 v_0 $.
Thick lines are the critical solutions which start at $v=v_0$
and $ v= \sigma_0 v_0$ at $\zeta=0$ and, in the case $\sigma_0=0$,
 asymptote  to zero velocity at infinity.
Dashed lines are the fast sonic lines, where $v =v_f$.
Dot-dashed line is  \Alfven  line, where $v$ equals \Alfven velocity.
At the point where fast sonic   line intersects \Alfven  line
the pressure is 0.
Characteristics intersect the fast sonic line vertically;
no physical solution can extend beyond such point.
Physical solutions start below the fast critical solution and,
in the magnetized case, above the low critical solution.
}
\label{topo}
\end{figure}

The  rhs of 
eq. (\ref{dg})  
 is equal to 0 only at  one point,
where the lhs of the eq. (\ref{dg}) is also $0$.
This
is a well studied critical sonic  point of the flow where $\beta=\beta_f$.
For a magnetized flow  the  
  rhs of 
eq. (\ref{dg}) become equal to 0   at two points:
a critical sonic point and another point located for a given
flow characteristic at  radii larger than the
sonic point. For even larger radii a sub-fastmagnetosonic
 mass-loaded flow starts to accelerate
- this behavior is  opposite  to the unmagnetized flow.
Acceleration of the flow proceeds very quickly;  the flow
reaches a point $\beta=\beta_f$ and $\partial_r \beta= \infty$.
A steady state flow cannot
exist beyond this limiting  radius.
These conclusion hold for both relativistic and non-relativistic
flows (considered in the appendix \ref{non-rel}.
The phase portrait of the mass loaded relativistic magnetized flows
is given in Fig.  \ref{toporel}, and non-relativistic in Fig.  \ref{topo}.

\subsection{Sonic point}

Sonic point occur when  both side rhs of the eq. (\ref{dg}) are
 equal to zero. It is possible to find the location of the 
point $\beta=\beta_f$ on the phase diagram $\beta-\zeta$.
 The general relations are complicated, yet simple 
approximations 
may be obtained in the limiting case of small and larger $\sigma_0$.
In the limit $\sigma_0 \ll 1$ the sonic point is located at a Lorentz factor
satisfying
\be
 3\,\left( 2- \Gamma  \right) \,\Gamma \,{\gamma_f }^3 +
 \left(4-  \Gamma  \right) {\gamma_f }^2 -
 \left( 2 + \Gamma  \right) \,  \gamma_f -3 =0.
\label{so}
\ee
 The position of the sonic point in radius is given by 
$\zeta_f =  f(\Gamma) = O(1)$, where $f(\Gamma)$ is a
complicated function of the order of unity.
For example for relativistic flows, $\Gamma =4/3$,
$\gamma_f= 1.09$, $\beta_f = 0.40$ and
$ \zeta_f =0.91$ while for non-relativistic flows, $\Gamma =5/3$,
$\gamma_f=  1.31 $, $\beta_f = 0.65$ and
$ \zeta_f =0.39$.

In the opposite limit  $\sigma_0 \gg 1$, we find
\be
\gamma_f^2 \simeq { 4- \Gamma \over 2(2- \Gamma )} \sigma_0
\ee
and $\zeta_f =  \tilde{f}(\Gamma)/\sigma_0^{5/2}$
where again $\tilde{f}$ is a
complicated function of the order of unity.
This gives
\ba &
 \gamma_f^2 = 2 \sigma_0, \, 
 \zeta_f= 1.19 \times  \sigma_0^{-5/2}
& \mbox{ for $\Gamma =4/3$ }
\nn &
\gamma_f^2 = {7 \over 2} \sigma_0,  
\zeta_f= 0.23 \times  \sigma_0^{-5/2}
& \mbox{ for $\Gamma =5/3$ }
\ea
Thus, low $\sigma$ flow experiences a shock transition when
the swept up mass flux becomes of the order of the 
luminosity of the central source, while 
high $\sigma$ flow experience shock transition at a distance
 $\sim \sigma_0^{5/6}$ closer to the source.
Location of the sonic points for two choices of $\sigma_0$ are shown in Fig.
 \ref{toporel}. 

\begin{figure}
\psfig{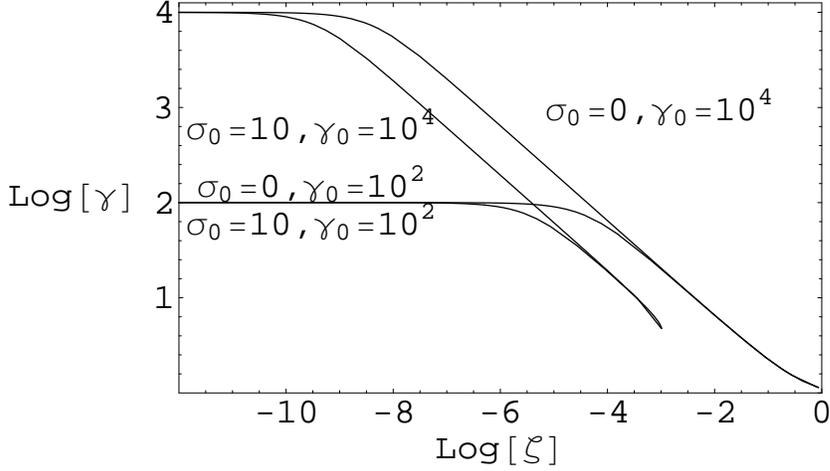}
\caption{Loading of supperrelativistic flows. Flows with higher initial
Loretz factors  and with higher magnetization parameters
 are affected by the  mass loading at smaller radii. Flows with the same
$\sigma_0$ approach the sonic point at weakly relativistic
velocities (not shown) along the similar
trajectories. Flows with  higher $\sigma_0$  experience a mass loaded
shock trasition at smaller radii than the flows with smaller $\sigma_0$.}
\label{fast-load}
\end{figure}

Since in the hyper-sonic regime $\gamma_0$ falls out of the equations, 
flows with different $\gamma_0$ but the same $\sigma_0$
  experience a shock
transition at the same location in $\zeta$ coordinate
(Fig. \ref{fast-load}).

\subsection{ Heavily loaded unmagnetized  winds}

For unmagnetized flows the effects of mass-loading  may 
 completely
dominate the dynamics of the flow at larger distances. This occurs
when the loaded mass flux becomes larger than the initial mass flux.
 In this limit 
$\dot{M} \gg \dot{E}_0,\, \dot{M}_0  $  and ${\cal E}=0$ it follows from the
eq. (\ref{dg}) that
 only non-relativistic strongly loaded flows
can extend to infinity.
Asymptotically
\be
\beta \sim r^{-  (5 \Gamma +1)/(\Gamma +1) } \propto r^{-7/2}
\ee
for $\Gamma =5/3$.

\subsection{Evolution of the  magnetization parameter $\sigma$}

Finally we consider evolution of the magnetization parameter $\sigma$. 
Restricting to $\Gamma=4/3$ we find
\be
{ \gamma^2  - \gamma_f^2 \over \sigma (1 + \sigma) }
 \, \partial_r  \sigma =-
\frac{ {\cal L } - {\cal F} \,\gamma  \gamma_f^2   }
  {  {\cal L } r  }
+ { \gamma ( \gamma-1 + 2 \beta_f^2 \gamma)  \gamma_f^2 \over 2 {\cal L } } \, R
\label{gf}
\ee
with $ \beta_f$ and $ \gamma_f$ given by eq. (\ref{pp}).
Thus effects of  mass loading are opposite to that of pressure.
In the absence of mass loading, pressure effects lead to a
$\sigma$ decrease for supersonic flows and a $\sigma$ increase
for subsonic flows. 
Mass loading contributes to an increase in 
$\sigma$   for supersonic flows and  decrease  for subsonic flows. 

In  mass loaded supersonic flows, the  pressure term always
dominates over the mass loading term in the eq. (\ref{gf}), so
$\sigma$ always increases, formally diverging at the point
$\beta = \beta_f$.

\section{Discussion}

In this paper we have investigated the dynamics of 
relativistic magnetized mass loaded flows. We  found that 
(i)
super-relativistic flows are effectively slowed down by mass loading;
(ii) subsonic magnetized flows subject to mass loading  have to {\it accelerate}
 beyond some radius and formally cannot reach infinity. We 
suggest that
internal instabilities would develop in the accelerating flow, destroying
the toroidal magnetic flux. Alternatively, the flow may become non-stationary.

To estimate whether mass loading is important in slowing
down pulsar winds we need to know the loading rate $R$ which depends on the
microphysics of the flow-ejecta coupling. To make a simple estimates we  
 assume that a considerable fraction
of the ISM neutrals (or  trapped ejecta particles)  with density 
$(1- \xi) n_{0}$, ionization fraction
$\xi$ and mass $m_p$
 are captured by the flow
on a characteristic time scale $\tau$:  $R  \sim n_0 m_p /\tau $.
Since in the Crab we do see filaments (i.e., they are not evaporated on a
dynamical time scale),
 for  static PWNs the
 age of a pulsar $T$
 may be considered as  a lower limit on $\tau \gg T$ and thus an 
 upper estimate of the rate $R$. From the eq. (\ref{r_d}) it follows that
the flow's initial Lorentz factor decreases  on a scale
\be
r\sim 10^{14} \left( {\dot{E}_0 \over 10^{36} {\rm erg} } \right)^{1/3} \,
\left( {(1-\xi)\,  n_0 \over 0.1 {\rm cm}^{-3}  } \right)^{-1/3} 
\left( {\tau \over  10^4 {\rm  yrs} } \right)^{1/3}  \,
\left( { \gamma_0 \over 10^6} \right)^{-2/3}  \,
\left( {1+ \sigma_0 \over 1001 } \right)^{-1/3}  \, {\rm cm}
\label{po}
\ee
We would like to stress that 
 this is a scale for a change of the initial Lorentz factor and not the
scale for the mass-loading-induced shock transition, which is much larger
 (see eq. (\ref{qp}).
Distance (\ref{po}) has a weak dependence on the  characteristic time scale
 $\tau$,
so that
 the deceleration of relativistic winds due to pick-up ions
 may be an important  factor affecting the evolution of pulsar winds.

To estimate the position where a mass-loaded  shock would occur we
assume that   pick-up rate is independent of radius. Then
the shock transition  occurs at
\be
r \sim \left( { 3 \dot{E}_0 \tau \over  4 \pi (1-\xi) n_0 m_p} \right)^{1/3}
\times
\begin{array}{cc}
1 &, \mbox{ for $\sigma_0 \ll 1$} \\
\sigma_0^{-5/6} &, \mbox{ for $\sigma_0  \gg  1$}
\end{array}
\sim 10^{17} \times {1 \over 1+ \sigma_0 ^{5/6} } {\rm cm}
\label{qp}
\ee
(for the same parameters as in eq. (\ref{po})).
Radius (\ref{qp}) also gives a typical location
where a flow becomes dominated by mass loading.
Radius (\ref{qp}) is of the order  of the shock distance in the Crab nebula.
Since it 
  is only a lower bound (since $\tau \gg T$) 
 we conclude  that a mass loaded 
 shock transition  is unlikely to occur in the static PWNs.
The shock observed in the Crab nebula nebula is  likely
an ordinary magnetohydrodynamical shock. 
On the other hand radius (\ref{qp}) is much smaller than the
radius of the Crab nebula. Thus  mass loading of the
subsonic pulsar wind should be very efficient.

Loading of supersonic flows greatly reduces the strength of the 
termination shock (the change
of the Lorentz factor of the flow), but 
not the kinetic energy flux through
it.  
This may affect the efficiency of particle acceleration and
PWN luminosity both in radio  and X-rays. 
Indeed, Williams et al. (1999) considered the strength of 
termination shocks
in mass loaded isothermal  flows and found that shocks become weaker.
They have also suggested that weakening of the shocks may contribute to the
radio quietness of some wind-blown bubbles.  Here we extend this possibility
to PWNs. Unfortunately, at  this point
our understanding of relativistic  shock acceleration  is not good enough to
argue whether weaker shocks (in terms of a change of a Lorentz factor)
are  less efficient at particle acceleration.

It is unlikely on numerical grounds that loading of  supersonic  flow
will slow it down to the critical point. If it still happens, then,
since 
for a broad variety of the loading profiles 
the critical point of the
flow is a focus and for a focus a smooth transition
from the super- to sub-fastmagnetosonic flow is not possible, 
 a flow would shock at a Mach number $>1$.
 For example, Galeev \& Khabibarkhanov (1990)
have argues that weakly magnetized non-relativistic  mass loaded
flows shock at a Mach number $M=2$. Similar analysis for relativistic flows
needs to be done as well.

We have shown that, contrary to naive expectations, 
 loading of  the relativistic shocked pulsar wind 
by particles from the filaments initially  would
result not in a slowing down of the flow, but in acceleration.
To see the dynamic reason for this strange behavior we first note that
even if a subsonic flow is weakly magnetized at the source, it becomes
strongly magnetized as it reaches its terminal velocity. 
The total  energy flux then
 consist of the Poynting flux $P = 4 \pi \gamma^2 \beta b^2$
and a particle flux $ \sim (\dot{M}_0 + \dot{M}) \beta^2/2$ (using the 
nonrelativistic expression for simplicity). 
The Poynting flux, subject to the requirement to transport
magnetic flux, $ \beta \gamma b = {\cal E}/2 \sqrt{\pi} r$,
is {\it inversely} proportional to the velocity:
$P = {\cal E}^2 /\beta$.  Mass loading increases particle energy flux,
so that  Poynting flux should decrease and  velocity should increase.

Acceleration of the wind would result in  development of instabilities
that would try to destroy the reason of the acceleration: the need to transport
the
magnetic flux. After the  instabilities
have developed and the reconnection destroyed magnetic flux the
flow will be allowed to decelerate, in line with the idea of Begelman (1998). 
This deceleration will be very quick, $\beta \sim r^{-7/2}$, allowing it
to match to the boundary of the PWN.

The other possible application of the present work 
relates to  the structure of the
ram pressure confined PWNs.
 The  pulsar motion through ISM constantly
brings a new neutral material in the wind.  The typical scale
in this case is the stand-off distance for the bow shock
\be
r_b = \sqrt{ {\dot{E}_0 \over 4 \pi c \xi n_0 m_p V_{NS}^2 }}
\ee
and a typical time  is $r_b/V_{NS}$ where $V_{NS}$ is the neutron star velocity
through ISM. Combining with (\ref{qp}) we find that the ratio of the
mass loaded induced shock transition to the stand-off distance
\be
{r \over r_b} = \left( { 3 \xi \over 1-\xi}  \right)^{1/3} \,
{1\over  (1+\sigma_0) } \,  \left( {V_{NS} \over c} \right)^{1/3}
\ee
which is typically smaller than unity.
Thus, mass loading is extremely important for the structure of the ram 
pressure
confined PWNs  (see also Bucciantini \& Bandiera 2001).

\begin{acknowledgements} 
I would like to thank Mikhail Medvedev, Vicky Kaspi, Elena Amato
 and Steve Balbus
for interesting and stimulating discussions.
\end{acknowledgements}

\appendix

\section{Mass loading of relativistic magnetized shocks}
\label{shock}

In this appendix we briefly consider mass loading of relativistic shocks.
Assume that at the shock  extra material with density $\Delta \rho$ in the
rest frame of the 
shocked plasma  has been
added to the flow. Then the
energy, momentum, mass conservation and  induction equations
are
\ba &&
\left[ \left( { \Gamma \over \Gamma -1} p + \rho + b^2 \right) \beta \gamma^2
\right] = \alpha
\nn &&
\left[ \left( { \Gamma \over \Gamma -1} p + \rho + b^2 \right) 
\beta^2  \gamma^2 + p +  b^2/2 \right] = 0
\nn &&
\left[  \rho \beta \gamma \right]=  \alpha
\nn &&
\left[  b \beta \gamma \right]= 0
\ea
where $\left[\right] $  implies a difference between shocked quantities 
(denoted with a subscript 2) and unshocked quantities 
(denoted below with a  subscript 1).
Introducing the ratio of the two four-velocities ${\cal N} = \beta_1 \gamma_1/
\beta_2 \gamma_2$, the ratio of the three-velocities 
${\cal R}=\beta_1/\beta_2$, upstream magnetization parameter $\sigma$
as a ratio of the magnetic energy density to particle  energy density
$h_1^2 = \sigma \left( \rho_1+ { \Gamma \over \Gamma -1} p_1 \right)$, 
and normalizing
we find 
\ba &&
\rho_1 \left\{ {\cal N}^2 
\left({{\cal R}-1 \over {\cal R}} 
( \beta_2^2 \gamma_2^2 (1+\sigma) - \sigma/\Gamma) -
{ (1+\sigma) \Gamma -1 \over {\cal R} \Gamma } + {\sigma \over 2} \right)+
\right.
\nn &&
\left.
 {\cal N} { \Gamma-1 \over  \Gamma} + 
+ {\sigma \over 2}  \right\}+
\nn &&
p_1  \left\{1+ {\Gamma \sigma \over 2(\Gamma-1)}+  {\cal N}^2 
\left(\sigma -{ \Gamma \sigma \over 2(\Gamma-1)}-{ 1 + \sigma \over {\cal R}}
 +{({\cal R} -1) \Gamma \over {\cal R} \Gamma}  (1+\sigma) \beta_2^2 \gamma_2^2
 \right) \right\}+
\nn &&
{ (\Gamma-1)(\gamma_2-1)- \beta_2^2 \gamma_2^2 \Gamma \over \beta_2 \gamma_2^2
 \Gamma} \, \alpha
\label{sa}
\ea
The mass loading term is positive definite.

For $\alpha=0$
 relation (\ref{sa}) reduces to 
 the known jump conditions for the relativistic magnetized
shock of arbitrary strength. For example for a strong (${\cal N} \gg 1$, 
$p_1 \ll \rho_1$)
 magnetized shock
this gives
\be
\beta_2= {1 \over 4 (1+ \sigma)} 
\left( 2 (\Gamma-1) + \Gamma \sigma \pm
\sqrt{( 2 (\Gamma-1) + \Gamma \sigma)^2 + 8 (2-\Gamma) \sigma (1+  \sigma)}
\right)
\ee
(this gives $\beta_2=\Gamma-1$ for $\sigma=0$).

Relation (\ref{sa}) shows that mass loading starts to affect the properties
of a strong shock when $ \alpha \sim  {\cal N}^2 \rho_1$.
A full investigation of the relation (\ref{sa}) is beyond the scope of this 
paper. 

\section{Non-relativistic mass loaded flows}
\label{non-rel}

Keeping in mind possible applications of the above results to
stellar outflows,
 below we briefly
consider non-relativistic  mass loaded magnetized winds.
The governing equations
\ba
&&
{1\over r^2} \partial_r \left[ r^2 \left( \rho v^2 /2 +
{\Gamma \over \Gamma -1} p + h^2 \right) v \right]=0
\nn &&
{1\over r^2} \partial_r
\left[r^2\left( \rho v^2 + h^2/2 \right) \right] + \partial_r p=0
\nn &&
{1\over r} \partial_r  \left[r h v  \right] =0
\nn &&
{1\over r^2 } \partial_r  \left[r^2 \rho  v  \right] =R,
\ea
may be simplified if one introduces energy, mass and magnetic  fluxes,
\be
{\cal L} = v \,\left( h^2 + {\Gamma \over \Gamma -1}
 \,p + \rho \, v ^2 /2  \right), \,
{\cal F} = v \,\rho , \,
{\cal K}=  v  \, h
\ee
which obey the equations
\be
\partial_r {\cal L} = - {2 {\cal L} \over r}, \, 
\partial_r {\cal F} = - {2 {\cal F} \over r} + R , \,
\partial_r {\cal K } = - {2 {\cal K} \over r}
\ee
with solutions
\be
{\cal F} = {\dot{M}_0 \over 4\,\pi \,r^2}  + {1 \over r^2} \int R\, r^2 dr, \,
{\cal L } = { \dot{E}_0 \over 4\,\pi \,r^2} , \,
{\cal K} =  { {\cal E} \over 2 \sqrt{\pi} r}
\ee

Introducing  a  fast magnetosonic  wave  phase velocity,
\be
v_f^2 = {h^2 \over \rho} + {\Gamma p \over \rho},
\ee
and
eliminating
$ {\cal K} $ in favor of $v_f$ we
get the equation for the  evolution of velocity:
\be
\left( v^2 - v_f^2 \right) {{\cal F} \over v}
\partial_r v =
{\Gamma -1 \over 2- \Gamma}\, { 2 {\cal L} - {\cal F} (v^2+ 2 v_f^2)
\over r} - {\Gamma -1  \over 2} R v^2
\label{as}
\ee

The first term on the rhs represents the effects of pressure
on the velocity; it is always positive
 since $p= (\Gamma -1)/(2 \Gamma v (2- \Gamma))
(2 {\cal L} - {\cal F} (v^2+ 2 v_f^2))>0$.
The negatively defined  second
term is proportional to the rate of mass loading.

When $R=0$ the
 evolution of the magnetized outflows can be integrated analytically.
Treating velocity as an independent variable we find from the
eq. (\ref{as}):
\be
r \propto  v^{ (2- \Gamma )/(2 (\Gamma-1))} \,
\left( 2 \dot{E}_0 v -\dot{M}_0 v^3-2 {\cal E}^2 \right)^{-1/(2 (\Gamma-1))}
\label{l}
\ee
Which shows that a terminal velocity at $ r \rightarrow \infty$
is determined by the third order equation for $v$ which involves the
energy, mass and magnetic fluxes at the source.

It is  convenient to introduce  velocity $v_0$ (the terminal
velocity of the super-fastmagnetosonic flow) and $\sigma_0$ (magnetization
parameter) by the following relations
\ba &&
\dot{E}_0 ={ \dot{M}_0 v_0^2 \over  2 (1- \sigma_0) }
\sim { \dot{M}_0 v_0^2 \over  2},
\nn &&
{\cal E}^2 = {\dot{M}_0 v_0^3 \sigma_0 \over  2 (1- \sigma_0)}
\sim {\dot{M}_0 v_0^3 \sigma_0 \over  2}
\label{KK2n}
\ea

When mass loading is not important the
 fast sound speed expressed in terms of $v$ and $\sigma_0$ is
\be
v_f^2 ={ \Gamma -1 \over 2} \left( v_0^2 - v^2 \right)
+{ 2 -\Gamma \over 2} {v_0^3 \sigma_0 \over v}
\ee
For weak magnetization $\sigma_0 \ll 1$ the fast sonic flow with $v =v_f$
is located at
\be
v= \sqrt{ (\Gamma -1)/(\Gamma+1)} v_0= v_0 /2
\mbox{ for
$\Gamma=5/3$}.
\ee

From the eq. (\ref{l}) it follows that
unmagnetized supersonic flow reaches
a terminal velocity  $v_0$ (and zero pressure),
 while  subsonic   flow slows down as
$ v \sim 1/r^2 \rightarrow 0$ (and constant pressure).
Magnetized flows   at $r=\infty$ behave somewhat different:
 at $r=\infty$ they satisfy
\be
v^3 - v v_0^2 + v_0^3 \sigma_0=0
\label{q}
\ee
Eq. (\ref{q}) has two real solutions for $ \sigma_0 \ll 1$ corresponding to
super- and sub-fastmagnetosonic
branches.
For a small but finite magnetization parameter, the
super-fastmagnetosonic branch remains almost unchanged, reaching
$v \simeq v_0$ at $r= \infty$, while the sub-fastmagnetosonic
branch asymptotes to a constant velocity determined by total magnetic and
energy fluxes:
$v \sim v_0 \sigma_0 \equiv {\cal E}^2 /\dot{E}_0$ (for $\sigma_0 \ll 1$).
Thus,
 magnetized flows cannot slow down to zero velocity since they have to
transport magnetic flux.

 Both sub- and super-fastmagnetosonic branches
at large radii have
 $ \rho \sim r^{-2}, \, p \sim r^{-2 \Gamma}$ and $h \sim r^{-1}$,
so that the
 plasma $\beta$ parameter (inverse of magnetization)
decreases with radius $\beta = 2  p /h^2 \sim r^{-2 (\Gamma -1)}$.
Both  super- and sub-fastmagnetosonic branches
 expand to zero pressure  at infinity.
Flows  which  were
weakly magnetized at the source (large $\beta$)
 become strongly magnetized as they expand.
In both cases magnetized flow have zero
 internal  pressure
at infinity, so that the terminal fast sound speed is equal the
shear \Alfven wave velocity:
\be
v_A^2 = {B^2 \over 4 \pi \rho} = { v^3_0 \sigma_0 \over 2 v}
\sim \left\{
\begin{array}{cc}
{v_0^2 \sigma_0 \over 2}&\mbox{at the fast branch $v \sim v_0$}\\
{v_0^2  \over 2}&\mbox{at the slow  branch $v \sim v_0 \sigma_0$}
\end{array} \right.
\ee

Next we consider 
special points
of the flow.
The  rhs of the eq. (\ref{as})
 becomes zero at
\be
\left({\Gamma+1 \over 2 (\Gamma -1)} R\, r +
{1 \over r^2} \int R r^2 dr \right) { 4  \pi r^2 \over \dot{M}_0}=
-1 + {v_0 ^2 \over v^2} -
{v_0^3 \sigma_0 \over v^3}
\label{s}
\ee
where parameterization (\ref{KK2n}) was used.
This equation has solution if the maximum on the  rhs (which is reached
at $v = 3  v_0 \sigma_0 /2$) is larger than
0, which limits magnetization parameter to
 $\sigma_0 < 2/( 3\sqrt{3}) \simeq 0.384$.

The condition $v= v_f$, given by
\be
(1+\Gamma) \left( 1+ { 4  \pi \int R r^2 dr \over \dot{M}_0} \right)=
(\Gamma-1) { v_0^2 \over v^2} +
(\Gamma-2) { v_0^3 \sigma_0 \over v^3},
\label{bs}
\ee
can be
satisfied at a given radius only for one particular value of $v$.

To proceed further we assume that
mass loading has a power law dependence on radius: $R \sim r^{-n}$.
Eqns (\ref{s}) and  (\ref{bs}) then become
\ba &&
{ ( 5+ \Gamma -  n (\Gamma+1) ) \over
2 ( \Gamma -1)} \, \zeta = - 1+{v_0 ^2 \over v^2} -
{v_0^3 \sigma_0 \over v}
\nn &&
(1+\Gamma) (1+\zeta) \left({ v \over v _0} \right)^3
- (\Gamma -1) { v \over v _0} - (2 - \Gamma) \sigma_0=0
\label{d}
\ea
where  a ratio of the mass flux due to loading to  the
initial mass flux
\be
\zeta = { 4  \pi \int  R  r^2 dr \over   \dot{M}_0} =
{ 4  \pi R r^3 \over (3-n ) \dot{M}_0}
\label{d2}
\ee
has been introduced. Below we assume that $\zeta$ is an increasing
function of radius and that $n< (5+\Gamma)/(\Gamma+1)$.

\begin{figure}
\psfig{file=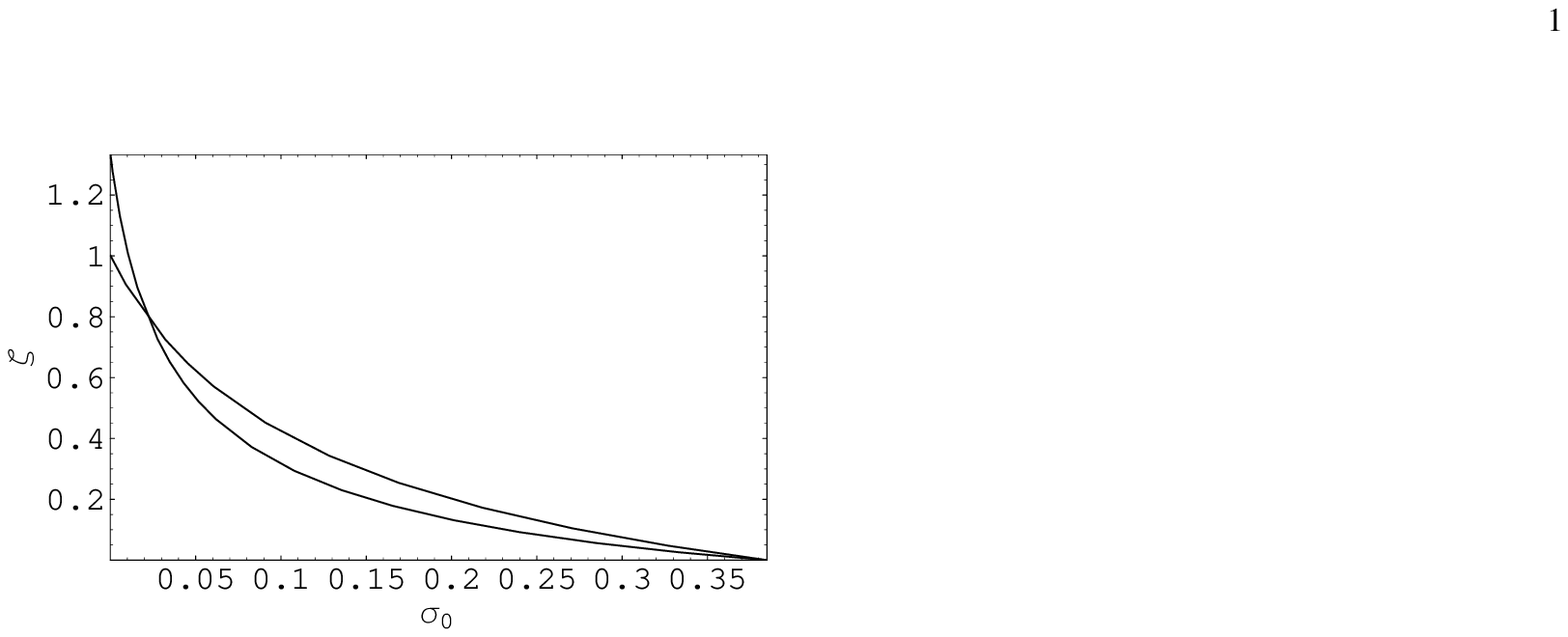,width=12cm}
\caption{Location of the fast sonic point for non-relativistic 
flows for  different values of $\sigma_0$.}
\label{sigma}
\end{figure}

 The system (\ref{d})  determines the location  $\zeta$ and velocity
at the fast  sonic  point. It can be resolve as an implicit
function of $v$:
\ba &&
{\sigma }_0 =
{\frac{v \,\left( \left( 3 - n \right) \,{v }^2\,
            {\left( 1 + \Gamma  \right) }^2 +
           \left( -1 + \Gamma  \right) \,
            \left( 1 + n - \left( 3 - n \right) \,\Gamma  \right) \,
            {{{v }_0}}^2 \right) }{\left( 3\,
            \left( 3 - \Gamma  \right) \,\Gamma  -
           n\,\left( 2 - \Gamma  \right) \,\left( 1 + \Gamma  \right)
           \right) \,{{{v }_0}}^3}}
\nn &&
\zeta= {\frac{2\,\left( -1 + \Gamma  \right) \,
         \left( -3\,{v }^2 + {{{v }_0}}^2 \right) }{{v }^2\,
         \left( 3\,\left( 3 - \Gamma  \right) \,\Gamma  -
           n\,\left( 2 - \Gamma  \right) \,\left( 1 + \Gamma  \right)
           \right) }}
\label{sg}
\ea
see Fig \ref{sigma}.
In the absence of magnetic field, $\sigma_0=0$,
relations (\ref{sg}) can be   resolved explicitly
for the location and velocity at the fast sonic point:
\ba &&
\zeta={ 4 \over \Gamma (3-n) -n-1}
\nn &&
v= \sqrt{ { (\Gamma -1) (\Gamma (3-n) -n-1) \over
(1+\Gamma)^2 (3-n)}} v_0
\ea
For non-vanishing  $\sigma_0$
the location of the sound point is pushed to smaller radii (and larger $v$),
 reaching
$\zeta=0$ at $ \sigma_0= 2/3 \sqrt{3}$ and $v= v_0/\sqrt{3}$
independent of $\Gamma$ and $n$.
In particular,
 for the simple case of  a
 homogeneous loading, $ n=0$, and $\Gamma=5/3$ the  fast sonic point is at
$
\zeta ={v_0^2 -3 v^2  /(5 v^2)}
, \,
\sigma_0 = {2  v / 5 v_0}  \,
\left( -1 + 8 { v^2 / v_0^2} \right)
$.

\newpage

\end{document}